# Simulation of complex phenomena in optical fibres

Jeremy Allington-Smith[1], Graham Murray and Ulrike Lemke[2]

Centre for Advanced Instrumentation, Durham University, Physics Dept, Rochester Building, South Rd, Durham DH1 3LE, UK

## Abstract

Optical fibres are essential for many types of highly-multiplexed and precision spectroscopy. The success of the new generation of multifibre instruments under construction to investigate fundamental problems in cosmology, such as the nature of dark energy, requires accurate modellisation of the fibre system to achieve their signal-to-noise goals. Despite their simple construction, fibres exhibit unexpected behaviour including non-conservation of Etendue (Focal Ratio Degradation; FRD) and modal noise. Furthermore, new fibre geometries (non-circular or tapered) have become available to improve the scrambling properties that, together with modal noise, limit the achievable SNR in precision spectroscopy. These issues have often been addressed by extensive tests on candidate fibres and their terminations but these are difficult and time-consuming. Modelling by ray-tracing and wave analysis is possible with commercial software packages but these do not address the more complex features, in particular FRD.

We use a phase-tracking ray-tracing method to provide a practical description of FRD derived from our previous experimental work on circular fibres and apply it to non-standard fibres. This allows the relationship between scrambling and FRD to be quantified for the first time. We find that scrambling primarily affects the shape of the near-field pattern but has negligible effect on the barycentre. FRD helps to homogenise the nearfield pattern but does not make it completely uniform. Fibres with polygonal cross-section improve scrambling without amplifying the FRD. Elliptical fibres, in conjunction with tapering, may offer an efficient means of image slicing to improve the product of resolving power and throughput but the result is sensitive to the details of illumination. We also investigated the performance of fibres close to the Limiting Numerical Aperture (LNA) since this may affect the uniformity of the SNR for some prime focus fibre instrumentation.

## 1 Introduction

Fibres have long been used for highly-multiplexed and precision spectroscopy. Despite their imperfections as optical components (internal absorption and non-conservation of Etendue) and the difficulty of manufacturing fibre systems, they remain the only practical option for multiplexed spectroscopy over large fields of view (e.g. 2dF - Colless et al. 2001; FMOS – Kimura et al. 2010). The importance of fibres will undoubtedly continue in the era of Extremely Large Telescopes due to the relative inaccessibility of their focal stations although this will inevitably mean longer fibres and new materials to maintain acceptable efficiencies, especially at shorter wavelengths. In the near-term they will continue to be used on the 4m-class telescopes on which such systems were initially developed, as these are re-engineered for widefield spectroscopy to address cosmological issues (e.g. BigBOSS, Schlegel et al. 2011). These include studies of dark energy via Baryon-acoustic oscillations (requiring spectral resolving power $R \approx 4000$ and multiplex gain, $G \approx 5,000$) and stellar archaeology to map out the history of galaxy assembly (requiring $R \approx 20,000$). The other main area where fibres will remain important is in precision spectroscopy such as that required for exo-planet studies to achieve radial velocity accuracies of a few cm/s at $R > 100,000$ (e.g. Wilken et al. 2010). Here it is the scrambling properties of the fibres and their stability which are the dominant issue in place of multiplex gain.

---

[1] j.r.allington-smith@durham.ac.uk
[2] Current address: Institut für Astrophysik, Friedrich-Hund-Platz 1, 37077 Göttingen, Germany

The fibres used in astronomy differ from those used in telecommunications mostly in that they have a relatively large aperture (60-300 μm) to couple light efficiently from the telescope to the spectrograph and so support many waveguide modes (10-500) at the focal ratios commonly used (F/3-F/7). Most of these multimode fibres (MMFs) have a step-index construction in which the core is surrounded by lower-index cladding so that light is totally-internally reflected. A range of materials are used but fused silica is very convenient for wavelengths up to 2.2μm. but beyond that, astronomers need alternative materials such as $ZrF_4$ and chalcogenides.

Despite the advantage that the flexibility of fibres simplifies the coupling of the sky to the spectrograph, allowing very high multiplex (e.g. G = 5000 for BigBOSS), there are a number of problems which must be carefully considered.

Etendue is not conserved by MMFs and is manifested by FRD (e.g. Poppett and Allington-Smith 2010 and references therein) which broadens the angular distribution emerging from fibre compared with the input. This requires enlargement of the spectrograph optics if throughput or spectral resolution is to be maintained.

Photometric errors may be caused by changes to fibres during tracking or between science and calibration observations. The main reason for this is modal noise caused by interference between the multiple modes in the fibres (Lemke et al. 2011, Lemke 2012 and references therein). This results in the formation of speckles which, when combined with vignetting by internal obstructions (such as the disperser stop) leads to variation in the throughput.

Scrambling, in which the output distribution is independent of the details of the illumination at the input, is important to precision spectroscopy. An error in the barycentre of the light transmitted to the fibre output will cause an error in both the radial velocity measured and in the spectral resolution as the width (and shape) of the PSF varies in the spectral direction.

Dealing with these effects requires not only high-quality fibres with low absorption over the band of interest, but a good understanding of FRD, scrambling and modal noise. This usually requires testing of candidate fibres. This can be a large task if one considers the need to optimise the fibre mounting by experimenting with different ferrule designs and materials, and the termination process itself, by experimenting with different methods of grinding and polishing or cleaving.

An alternative is to simulate some of these effects. This is possible by modal - analysis or by ray-tracing. Snyder and Love (1983) and Heacox (1987) provide a rigorous theoretical framework based largely on modal analysis which Watson and Terry (1987) and Barden et al. (1993) were able to use in their own experimental studies. Ray-tracing can be done using a commercial software packages but these do not generally take account of the more complex phenomena such as FRD. In this case, the model must not only provide a mechanism to simulate the modal diffusion process postulated by Gloge (1971), but also the lack of the dependence on length which that model predicts; which has recently been addressed by Poppett and Allington-Smith (2009). Since it is widely believed that scrambling requires some FRD to be present, without the ability to simulate FRD, it is unlikely that any simulation program can make useful predictions of scrambling issues.

It is important to realise that scrambling is not just a problem for precision spectroscopy but also affects highly-multiplexed spectroscopy which may, for practical reasons, include connectors in the fibre run from the on-telescope pickoff to the spectrographs. In budgeting for throughput in such complicated systems it is necessary to model the distribution of light at each interface so that tolerances on the angle and displacement of the mating parts of the connectors can be assessed. This is critically dependent on the shape of the nearfield and farfield light distributions as we will demonstrate. The success of these projects, for example, BigBOSS, is predicated on strict control of these error budgets.

In this paper, we present a model which allows us to simulate FRD and other non-standard features of optical fibres. After a description of the model in §2, we show how it reproduces some basic aspects of fibre behaviour in §3. In §4, we show how FRD can be implemented and how it affects scrambling. In §5, we discuss other problems of fibre use including the Limiting Numerical Aperture (LNA). In §6, we widen the topic to include non-circular fibres and show how the model allows us to predict their performance.

## 2  Modelling fibres

### 2.1  Optical interfaces

We can consider three types of optical configuration in which fibres are commonly used.

**Configuration A:** For traditional fibre multi-object spectrographs fed from a telescope with a fairly fast focal ratio (<f/8), the fibre can be placed directly at the focal surface and arranged into a pseudoslit which matches the curvature of the focal surface and orients the fibres to the exit pupil of the telescope. Therefore the fibre nearfield, comprising partly-scrambled images of the sky, is projected on the detector while the disperser is illuminated by the fibre far-field.

For other types of spectrograph, especially integral field spectrographs, the fibre input will may be coupled via a lenslet array which images the telescope exit pupil on the fibre core. When used with magnifying fore-optics, this can eliminate the deadspace between cores and adapt the input focal ratio to minimise FRD. The two variants are as follows.

**Configuration B:** The pseudoslit includes a linear lenslet array to convert back to the native focal ratio of the telescope. Therefore the fibre far-field, comprising doubly-scrambled images of the sky is projected onto the detector, while the disperser is illuminated by the reimaged fibre near-field comprising scrambled images of the telescope pupil.

**Configuration C:** The pseudoslit consists of the bare ends of the fibres, possibly interfaced to a glass plate to reduce scattering losses. Therefore the fibre near-field, comprising scrambled images of the telescope exit pupil, are projected on the detector while the disperser is illuminated by the fibre far-field comprising scrambled images of the sky

It can be seen that the optimum configuration depends on the accuracy required in the near- or farfield and the impact of non-uniform illumination of the grating.

## 2.2 Types of ray

Rays can be characterized as either meridional or skew. The former pass only through a plane (meridian) containing the axis of the fibre (Figure 1). Since, in the absence of perturbations (which will be added later), the rays have no transverse component, they will stay within the meridian into which they were launched. However, skew rays which are not constrained to a meridian can strike the wall of the fibre at an angle with components both parallel to the meridian and transverse to it. Thus these rays will tend to follow a helical path through the fibre which may entirely avoid the axis (see Heacox 1987, Watson and Terry 1995) .

## 2.3 Geometry

The coordinate system is an orthonormal set with $z$ measured along the fibre axis in units of the fibre radius. Thus $z = 1000$ implies a length of $L = 50$mm for a fibre of diameter $2R = 100$μm. The ray direction is given by a vector $\underline{U}$ with angles $\alpha = \arctan(U_x/U_z)$ and $\beta = \arctan(U_y/U_z)$ with respect to the $z$-axis. Positions are given by a position vector $\underline{P}$. The geometry of the fibre is shown in Figure 1. The illuminated spot on the fibre core is defined by the azimuth angle of its centre, $\psi$, its radial displacement, $r$ and the spot diameter, $s$.

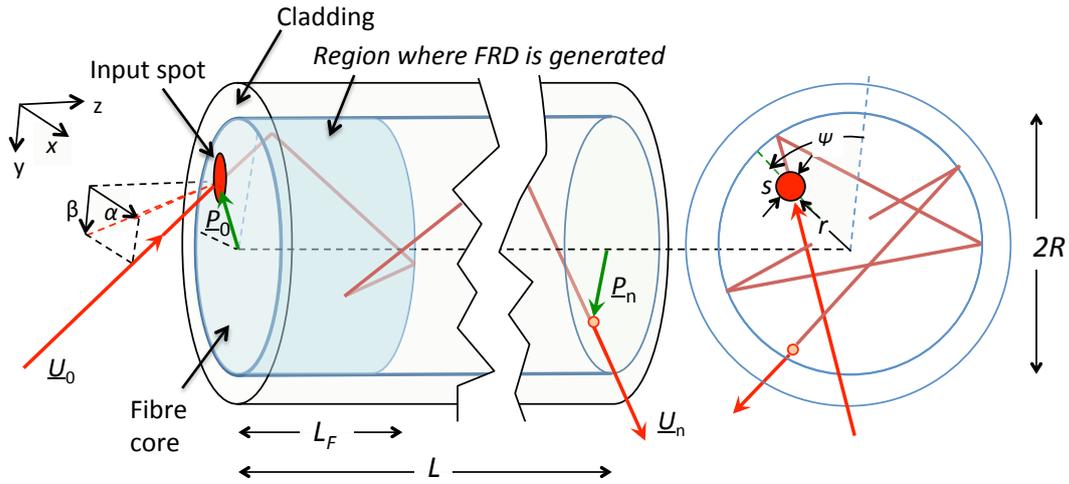

*Figure 1; Geometry of simulation. The sketch shows how a light ray enter the fibre along vector $\underline{U}_0$ via a defined spot and propagates along the fibre until it exits long vector $\underline{U}_n$ after n reflections. In this case, the rays are skew since it is not restricted to a meridian such as that defined by the position vector $\underline{P}_0$ which represents the initial intersection of the ray with the fibre input face. The shaded region indicates where FRD is generated (see §4).*

**Ray production.** Rays are launched from a specified region of the input face of the fibre with a range of angles characterized by a top-hat function having uniform surface brightness within a specified range of angle defined by either the input focal ratio, offset from the direction of the fibre axis as required, or the input angle of a collimated beam. The set of rays are thus distributed with each of the 4 degrees of freedom chosen at random following a given seed.

**Macro-bending.** It is necessary to include long-scale curvature into the fibres for two reasons: firstly to provide a more realistic simulation and secondly to avoid pathological singularities which might occur otherwise (e,g, rays launched at very small angles into a perfectly straight fibre will never strike the fibre wall). Macrobending is simulated by perturbing the angles at which rays strike the fibre wall to simulate the effect of the fibre being bent into half a cycle of a cosine function with an amplitude specified by curvature $C = R/r_C$. where $r_C$ is the radius of curvature

**Removal of super-critical rays.** The angular perturbations introduced may cause the angle at which the ray strikes the core/cladding interface, $\xi$, to fall below the critical angle, measured to the local normal to the fibre wall:

$$\xi_C = \frac{\pi}{2} - \arcsin\Theta$$

where Θ is the limiting numerical aperture (LNA) of the fibre which depends on the relative refractive indices of the fibre core and cladding[3]. Once the condition $\xi < \xi_C$ is detected, the ray is terminated. Although the critical angle is related to the LNA of the fibre, this only applies to meridional rays. As we will show, skew rays can still travel through the fibre at higher angles depending on the geometry of illumination. The simulation of Focal Ratio Degradation is discussed in §4.1.

**Phase-tracking.** The program has been designed to keep track of the total path length of each ray. This means that the effect of coherence can be simulated as a means to address (to some level of approximation) modal noise.

## 2.4 Implementation

The ray-tracing program was implemented in *MathCad*. The computational procedure is similar to that used for production of photorealistic scenes for engineering and gaming. The key tasks are, firstly, to find the point of intersection of the ray with the boundary of the 3D object representing the fibre; and then to determine the path of the reflected ray relative to the normal to the surface of the 3D shape. This procedure is repeated until the end of the fibre is reached, where the fibre exit angle and the position of the ray intercept with the fibre end-face is calculated. The location of the intercept and the normal vector may either be calculated analytically or found numerically. The latter is more practical for complex shapes (e.g. anamorphically tapered fibres) but makes the program execute more slowly.

This program produces visual representations of individual rays via isometric views of the ray trajectories as well as the final positions of the ray ($P_x$, $P_y$) as it passed through a plane representing the fibre endface and the corresponding angular vector ($U_x$, $U_y$). This ray-tracing engine was then used by another program to generate the originating rays, with the specified range of position and angles chosen at random. The results were visualised as a scatter plot of the ray coordinates relative to the fibre aperture and the angular extent of the input ray angles. These coordinates were used to generate the RMS and mean of $P_x$, $P_y$, $U_x$, $U_y$. In addition, these Cartesian coordinates were transformed into a polar basis and plotted to reveal, for example, the amount of FRD via the width of the ring formed in the farfield when collimated light is input at a specified angle.

For maximum clarity, the number of traced rays shown in the figures is 1000, sufficient to give a clear visual distinction between essentially uniform and non-uniform distributions, avoiding the sampling issues which affect binned data. Quantitative results for a given configuration were obtained from 10 independent spot diagrams to improve statistical significance. Unless stated otherwise the fibre length is 1000*R*.

The simplest cases to simulate are described below. These may be compared to the experimental result of Watson and Terry (1995), Barden et al. (1993), Avila and Singh (2008) and Avila et al. (2010) and the theoretical exposition of Heacox (1987).

---

[3] The refractive index of the fibre core does not enter into this expression because, for simplicity, we assume that the rays propagate in a medium with unit refractive index.

# 3 Simple behaviour

## 3.1 Illumination by collimated light

A fixed angle of illumination produces a ring in the farfield which is uniform in azimuthal angle and uniform in the near-field distribution (*Figure 2*) as expected. The output rays preserve the radial component of the input angle since there is no mechanism to change it, but the azimuth angle is randomised as skew rays populate all azimuth angles. The near-field pattern is determined by the location where rays intercept the output face. Since different rays entering different parts of the input face will follow different paths with different distances between successive reflections, the arrival positions will be random.

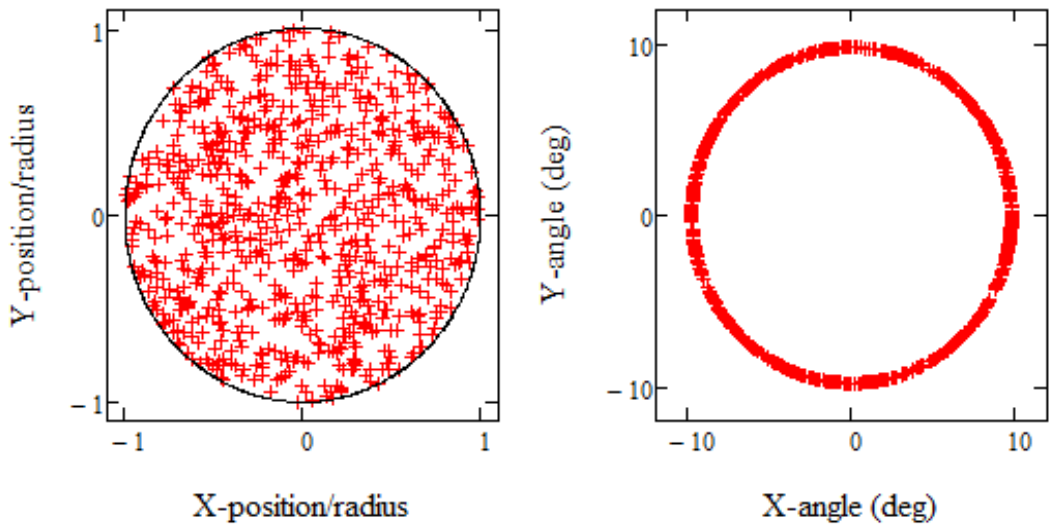

*Figure 2: Spot diagrams for nearfield (left) and farfield (right) for illumination by a collimated beam injected at angle 10 deg to the fibre axis and filling the input face of the fibre. The outer circle on the left indicates the fibre core (units are multiples of the core radius).*

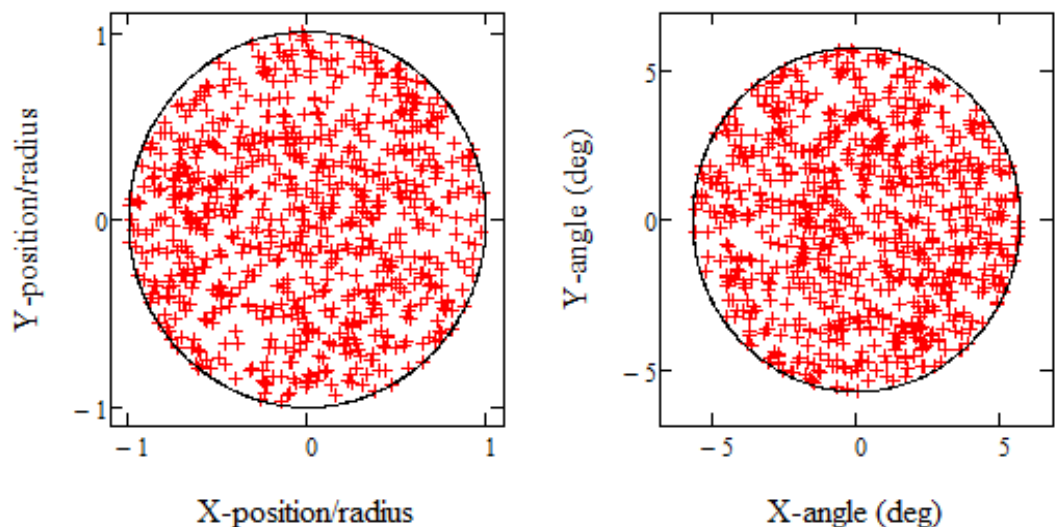

*Figure 3: Illumination with a filled f/5 cone over the entire input face.*

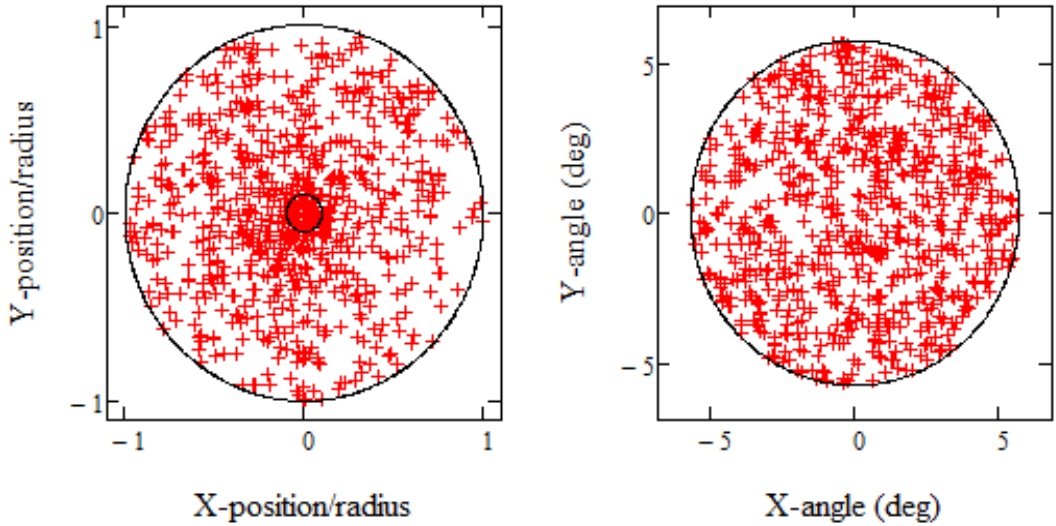

*Figure 4: Same as previous figure but for a centrally-located spot (indicated by the small circle) at F/5. The outer circle on the left indicates the fibre core (units are multiples of the core radius). The circle on the right indicates the angular extent of the cone (of uniform surface brightness) illuminating the fibre.*

### 3.2 Illumination by a filled cone of light

Illumination by a filled cone of illumination produces radically different outputs depending on the size and location of the illuminated region in the input face. If it covers most of the fibre face, the nearfield pattern is uniform (*Figure 3*) but if the illuminated region is a small spot on axis, the nearfield pattern is sharply peaked at the centre (*Figure 4*). In either case, the farfield pattern is uniform with a clean cutoff at the maximum angle of the input cone.

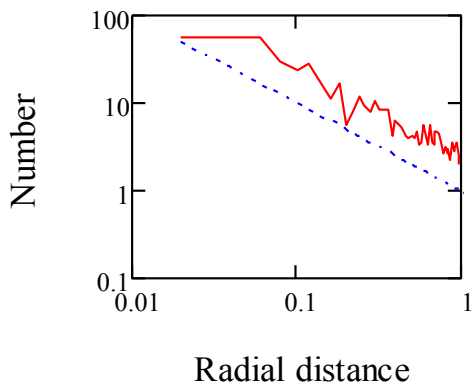

*Figure 5: The nearfield distribution with radius of the traced rays in the preceding figure (red). The dotted line has a slope of -1 predicted in the text.*

This behaviour is confirmed by Watson and Terry, Barden et al., Avila and Singh and Avila et al., and explained by Heacox and Watson and Terry. Since rays launched close to the fibre axis will be predominantly meridional, we expect that they will be uniformly distributed within each meridian. Therefore the position of the exiting ray within each meridian will be random so the ray density, $\rho$, will be independent of radius. Thus $d\rho/dr$ = constant, but the surface brightness $S = d\rho/dA = (d\rho/dr)(dr/dA) \propto 1/r$.

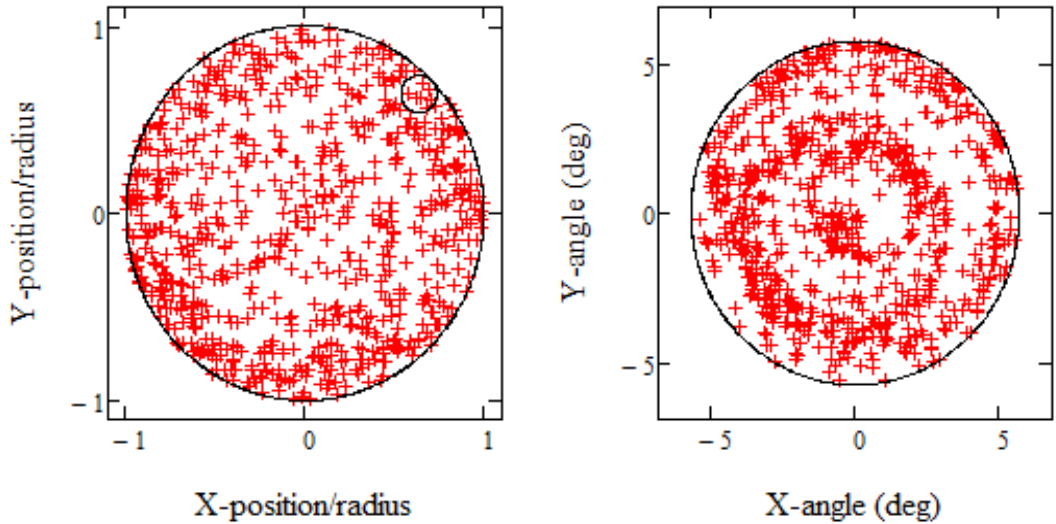

*Figure 6: Same as Fig. 2 but with spot offset to the edge of the fibre face (small circle) and length 100R. Note the lack of homogeneity in the farfield.*

*Figure 5* shows that the surface brightness is indeed proportional to $1/r$. In fact this becomes uniform near the centre at a radius that depends on the size and position of the spot. As the spot moves off-axis, the nearfield pattern will become more like that in *Figure 3* because the rays will become predominantly skew so that they will no longer be confined to a meridian and can therefore produce a scatter in arrival positions over the entire face. In the extreme case of a small spot at the edge of the fibre input face (*Figure 6*) the skew rays dominate the picture to the extent that most rays are confined to the outer parts of the fibre causing a nearfield distribution that is edge-brightened.

In this figure, we also see evidence for incomplete scrambling in the farfield. This is also confirmed qualitatively by Avila et al (2010) who recorded a sequence of bright and dark rings as the illumination spot moved across the fibre face. This example is for a short fibre of length only 100*R*, i.e. 5(10)mm. If we increase the length a hundredfold the farfield distribution become homogenous but the nearfield pattern remains edge-brightened (*Figure 7*). See §5 for the effect that this has on scrambling.

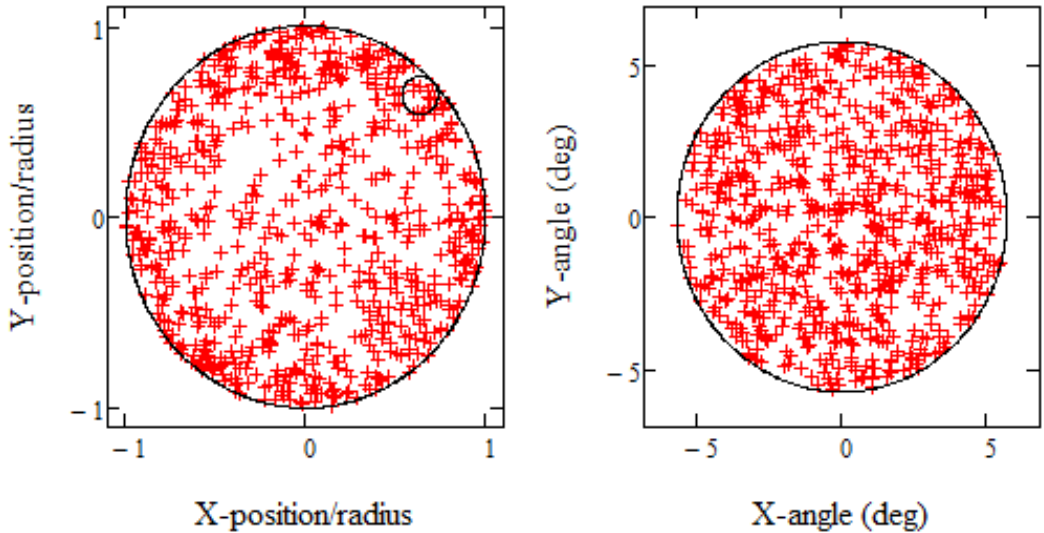

*Figure 7: Same as previous figure but with a fibre 100 times longer. The farfield is now quite homogenous although the near-field is significantly edge-brightened.*

## 3.3 Macrobending

Although it is well known that severe bending of fibres affects the distribution at exit, it is useful to quantify this even for very strong curvature because of the impact that the minimum fibre bend radius has on the design of instruments, especially those that employ many actuators packed into a small space. Clearly, it is important to know the smallest bend radius that gives adequate throughput.

The simulated effect of macrobends is shown in Figure 8. This shows the results of simulations with a fixed, strong curvature ($C$ = 0.01) as a function of fibre length when the fibre core is illuminated with a filled cone of F/5. Rays at larger angles are lost progressively as the length of fibre increases from 1,000 to 30,000 fibre radii leading to a reduction in throughput and a reduction of the angular extent of the farfield which would lead to an underfilling of the disperser for Configurations A and C leading to loss of spectral performance. Light is lost due to the rays exceeding the critical angle – an effect which accumulates as the light passes though the fibre. Note that this is one effect where the scrambling (in the farfield) gets worse with increasing fibre length!

It is clear that the macrobending, if very tight ($C$=0.01 implies a bend radius of 100 times the fibre radius. i.e. 5(10)mm for $R$=50(100)μm) severely affects both throughput and farfield illumination but a bend radius of 50(100)mm has little effect (Figure 9). We find that varying the amount of FRD also has little effect .

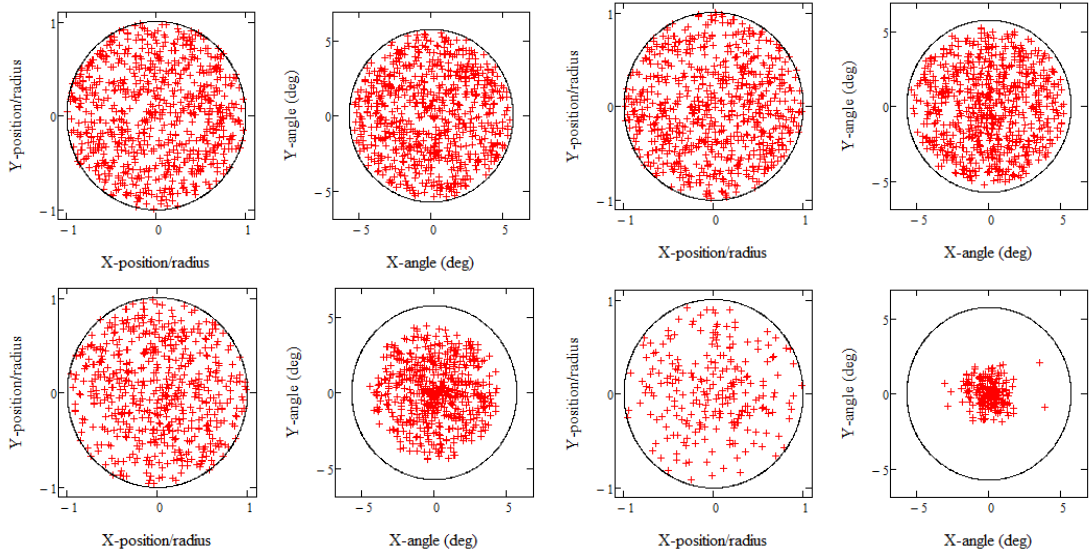

*Figure 8: Loss of light at the edge of the far-field due to severe macrobending (C=0.01) for lengths of 1,000, 3,000, 10,000 and 30,000 times the fibre radius. From top-left to bottom-right. The corresponding throughputs are 96, 92%, 71% and 31% (all ±1%).*

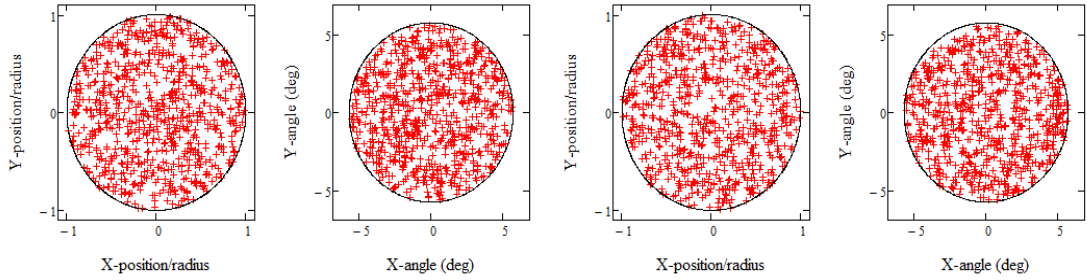

*Figure 9: Same as the previous figure but for fibres of length 1,000 and 30,000 times the fibre radius only, with reduced curvature (C=0.001). Throughputs are 100% and 99% respectively.*

## 4 Focal Ratio Degradation

### 4.1 Simulation by ray perturbations

FRD is well described by the Gloge (1971) model. This postulates the existence of scattering sites distributed within the fibre which scatter the light into higher and lower modes; i.e. within the context of the ray model, to larger or smaller angles. Therefore it seems natural to simulate this modal diffusion by random angular perturbation in ray trajectories. However, the Gloge model predicts that FRD should scale linearly with fibre length $L$, which is not found in most experimental work. This has been explained by Poppett and Allington-Smith (2010) as being due to an end-effect where the scattering is concentrated at the ends of the fibres where they are terminated. This might be due to mechanical and thermal stress induced during termination (usually including grinding and polishing). Indeed, the ray-tracing model did exhibit the tendency for the FRD to increase with fibre length when the scattering was applied along the full length of the fibre. Consequently, we simulated the end-effect by restricting the angle randomization to a specified length at the fibre input.

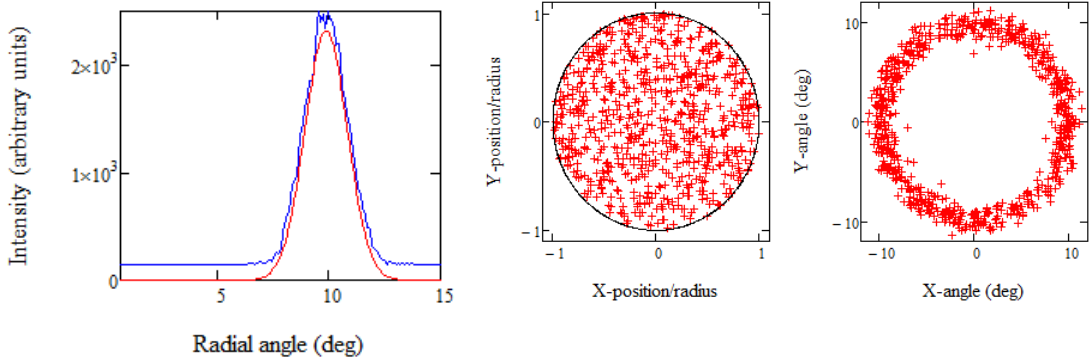

Figure 10: Calibration of FRD. Left - the radial profile of the farfield obtained from the simulated data shown at the right is well fit by the Gloge model (left). The model (red) and simulated data (blue) are offset vertically for clarity.

## 4.2 Calibration of FRD

Although the program reproduces the essential features of FRD, there is no natural link between the parameter that describes FRD in Gloge's model, $d_0$, which represents the linear density of scattering centres, and the parameter, $\Delta\theta_F$, in our numerical model, which determines the amplitudes of angular perturbations in the ray-tracing model. Therefore we perform a calibration whereby we vary the perturbation amplitude until it produces a realistic amount of FRD. To fit the observations, it is also necessary to ensure that the amount of FRD is constant with fibre length and does not increase as naïve application of the model would predict. We then note the required perturbation amplitude and fit the predicted data to the Gloge model and note the value of $Ld_0$ required. These parameters are degenerate because of the observed lack of length dependency (see above).

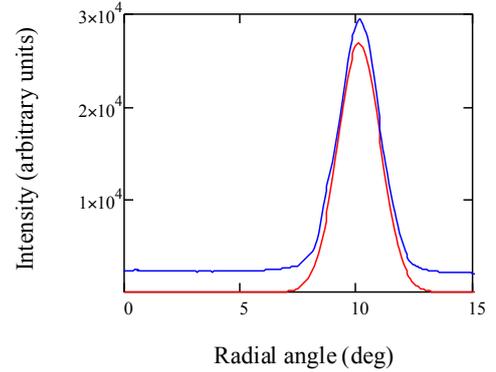

Figure 11: Same as for the left-hand plot of the previous figure but for measured data for a typical fibre. Error bars on the data are too small to be plotted.

To account for the lack of length-dependency, we restrict the scattering process to a length $L_F$ at the fibre input (see **Figure 1**). This value is chosen to be similar to the distance along the fibre which may be subjected to stress frozen in during manufacture. In particular, it represents the depth to which the fibre adhesive may wick up in the gap between the fibre outer diameter and the inner diameter of the ferrule. Although it is likely that both ends will be subject to the same type of termination, we can combine these two lengths into one without loss of generality. In any case, the choice of $L_F$ is not critical provided that the combined effect of $\Delta\theta_F$ and $L_F$ correctly provides the amount of (length-independent) FRD measured for the fibre under test, and that $L_F$ is smaller than the shortest fibre length which might be actually tested.

*Figure 10* shows that the simulated output is well fit by the Gloge model indicating that it provides a good phenomenological description of FRD. In this example, *Δθ_F = 0.020* produces a ring with *FWHM = 2.2±0.05* deg and Gloge parameter *$L_F d_0$ = 280±10*. This example was chosen to reproduce results obtained with a real fibre (Polymicro FBP) with the same core diameter, 120µm and LNA = 0.22. Tests on fibres of this type produce a range of ring FWHM 1.1 – 2.3, so we have used a fibre with FRD near the upper limit to provide a stringent test.

The fibre was illuminated over the full endface by a low-power HeNe laser at an angle of 10 deg to the fibre axis. The resulting farfield distribution was recorded by a CCD. The centroid of the distribution was estimated and the ring summed in azimuth around this point. The ring width is 2.1±0.05 deg FWHM (*Figure 11*). The distribution is well fit by the Gloge model with *$L_F d_0$ = 290±10*. This is in good agreement with the simulation indicating that the effect of FRD can be simulated accurately with the appropriate calibration between $Δθ_F$ and $L_F d_0$ for fixed $L_F$.

To demonstrate the lack of length dependence, as seen in Poppett and Allington-Smith, we repeated the simulation shown in Figure 9, which was done for length 1000*R*, for lengths 100*R* and 10,000*R*. Within the measurement errors, there was no difference in the ring thickness.

*Figure 12* shows that the ring thickness is roughly constant with the angle of the input beam for angles exceeding 5 deg. At smaller angles, the fitting of the ring profile is complicated by light that appears on the other side of the centroid depending on the amount of FRD. Table 1 shows results from several fibres of the same type as the single fibre tested above for input angle ±6.5 deg. This indicates the variability between fibres and the repeatability of the fibre measurements. Analysis of the results suggests that the experimental error is 0.1 deg and that the intrinsic variability of the fibre FWHM in this sample is 0.53 deg. In this paper we refer to "typical FRD" as that obtained with $Δθ_F$ = 0.015, i.e. that which produces a ring thickness of 1.6 deg FWHM which is close to the mean value in Table 1.

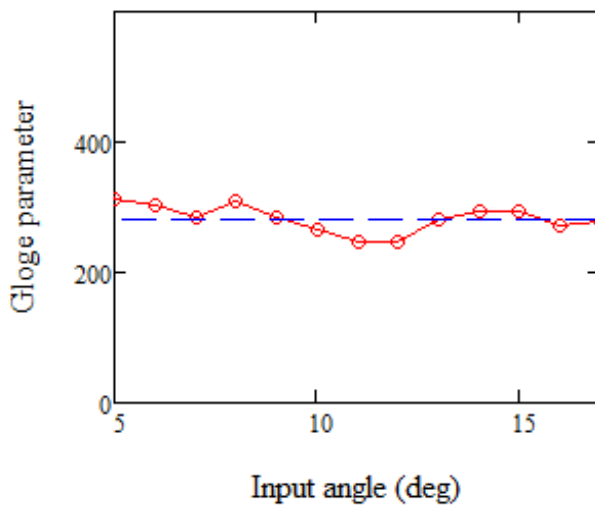

*Figure 12: Value of the Gloge parameter, $L_F d_0$, obtained for the tested fibre with different input angles of the collimated beam. The uncertainty in the Gloge parameter is estimated as ±30*

Table 1: FRD results for a set of fibres expressed as the FWHM of the ring thickness for equal and opposite input angles of 6.25 deg. Also shown is the mean of the two values and their dfference.

| Fibre ID | +6.25 deg FWHM | error | -6.25 deg FWHM | error | Mean FWHM | error | Difference FWHM |
|---|---|---|---|---|---|---|---|
| A | 1.71 | 0.11 | 1.74 | 0.11 | 1.73 | 0.11 | -0.03 |
| B | 2.32 | 0.27 | 2.51 | 0.17 | 2.42 | 0.22 | -0.19 |
| C | 1.11 | 0.07 | 1.37 | 0.03 | 1.24 | 0.05 | -0.26 |
| D | 2.05 | 0.05 | 2.30 | 0.19 | 2.18 | 0.12 | -0.25 |
| E | 1.14 | 0.02 | 1.20 | 0.04 | 1.17 | 0.03 | -0.06 |
| F | 0.92 | 0.03 | 0.82 | 0.09 | 0.87 | 0.06 | 0.10 |
| G | 1.20 | 0.10 | 1.08 | 0.17 | 1.14 | 0.14 | 0.12 |
| H | 1.10 | 0.06 | 1.31 | 0.12 | 1.21 | 0.09 | -0.21 |
| Mean | 1.44 | 0.09 | 1.54 | 0.12 | 1.49 | 0.10 | -0.10 |

## 4.3 Surface scattering

As highlighted by Haynes et al. (2008), scattering at the fibre endface may produce additional beam-spreading which might contribute to FRD. Consequently we have included this effect in our model.

The scattering is presumed to originate from the propagation of rays through the fibre face if that surface is considered to be made up of large number of facets randomly oriented over a range of angles with respect to the fibre axis, $\theta$. The probability of a ray encountering such a facet is $p(\theta) \propto \cos^2\theta$. If for simplicity, we assume that the rays are perturbed by an amount proportional to $\theta$, then a coordinate in the plane of the farfield image, $U_r = \tan\theta$, will follow a Cauchy (Lorentzian) distribution:

$$p(U_r) \propto \left[1 + \left(\frac{U_r}{w}\right)^2\right]^{-1}$$

where $w$ is the FWHM. This is in contrast to the farfield Gaussian distribution produced by the description of FRD by Gloge in which the scattering takes place throughout certain volumes of the fibre. It is also likely to differ from the Gloge model in being directly caused by irregularities in the fibre surface rather than sub-surface stress whose effects are be partly mediated by optical activity. However, both mechanisms share an independence of fibre length since they occur only at or near the fibre ends.

However, in practice, we do not often find it necessary to include this effect to understand the output light profile. The example data shown in *Figure 11* is well-fit by the Gloge model which implies that the surface scattering effect is minor in this case. This fibre was polished using standard techniques with media containing sub-micron particles which gives a very smooth finish. A common way of terminating fibres is to immerse them to a smooth glass plate after polishing using an index-matching gel. This reduces the surface scattering even further but was not employed in this case.

Likewise the set of fibres described above also showed little evidence for a Lorentzian distribution. Although some scattered light could be detected at low levels, this appeared to be due to stray light in the apparatus.

For these reasons, although we have implemented surface scattering in the model by generating random perturbations at the fibre exit following a Lorentzian distribution scaled by *w*, we have not included surface scattering in our results.

## 5  Scrambling

The effect of imperfect scrambling has been investigated by simulating the effect of placing a spot smaller than the fibre core at various radial positions. The cases considered are for a spot radius *s* = 0.1*R* and 0.3*R* offset by various amounts from the fibre axis. Typical FRD is included in the simulations.

It was found that the centroid of the light distribution had no dependence on the position of the input spot even for fibres as short as 300*R* irrespective of the presence of (simulated) FRD. However there was a noticeable change in the shape of the light distribution, parameterized by the RMS, as shown in Figure 13.

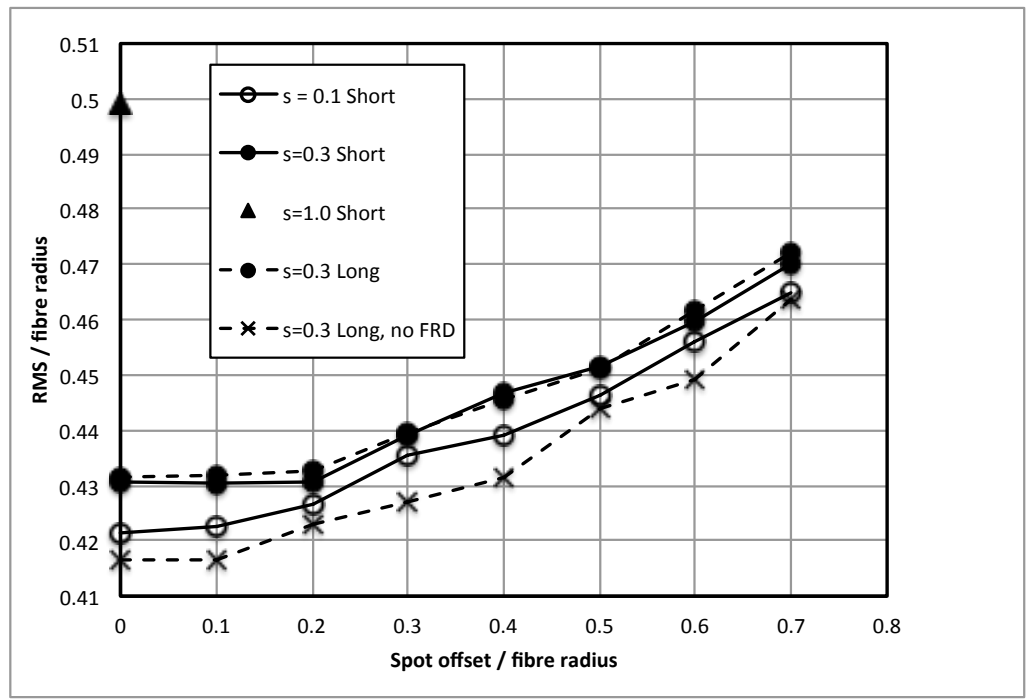

*Figure 13: RMS of nearfield light distribution as a function of the offset of the input light spot from the centre. The cases shown are explained in the text of §5.*

First consider the short fibres (length $300R$) for illuminated spot radii of 10% and 30% of the fibre radius (solid lines). When the spot is on axis it behaves rather like the case seen in *Figure 4* where the surface brightness scales inversely with radius to give a sharp central spike truncated by the size of the spot. This naturally produces a fairly low RMS. Physically, the peak is caused by the predominance of meridional rays over skew rays, an effect which diminishes with increasing spot size and increasing offset. Thus we find that the RMS increases as the distribution loses its peak as the offset increases and also as the spot size increases. For reference, a perfectly uniform distribution over the full face has RMS = $0.5R$.

Increasing the fibre length tenfold to $3000R$ makes no difference (dashed line with filled circle). Removing the effect of FRD (crosses) leads to a more peaked distribution indicating that FRD does indeed aid scrambling although not enough to totally homogenise the distribution of light on the input face.

# 6 Behaviour near the limiting Numerical Aperture

The limiting numerical aperture is conventionally used to describe the maximum width of the beam that can be fed into a fibre. It is determined by the structure of the fibres, specifically the relative refractive indices of the core and cladding as required for total internal reflection (see above). However, this derivation is only valid for meridional rays, not for skew rays. By illuminating the fibre in different ways the relative proportions of meridional to skew rays can be altered with the results that the limitation imposed by the LNA can be defeated. This also means that, for a practical configuration in which the fibre is illuminated with a range of angles such as in a filled cone, the LNA is not a hard limit so that the throughput varies with input focal ratio near the LNA rather then being suddenly truncated when the LNA is reached.

To demonstrate this using the ray-tracing model, we have illuminated the fibre input via a small spot at one specified angle close to the LNA whose azimuth direction is altered from 0 to 90 deg. (We refer to the difference in the azimuth of the illumination spot and the azimuth of the input ray direction as the illumination azimuth; ***Figure 14***). This shows that configurations where the rays strike the core-cladding interface obliquely produce a spiralling of rays which meet the fibre wall at a relatively shallow angle. However when light strikes the fibre wall with one component of motion radial to the wall, the ray is more likely to exceed the critical angle and not be transmitted further.

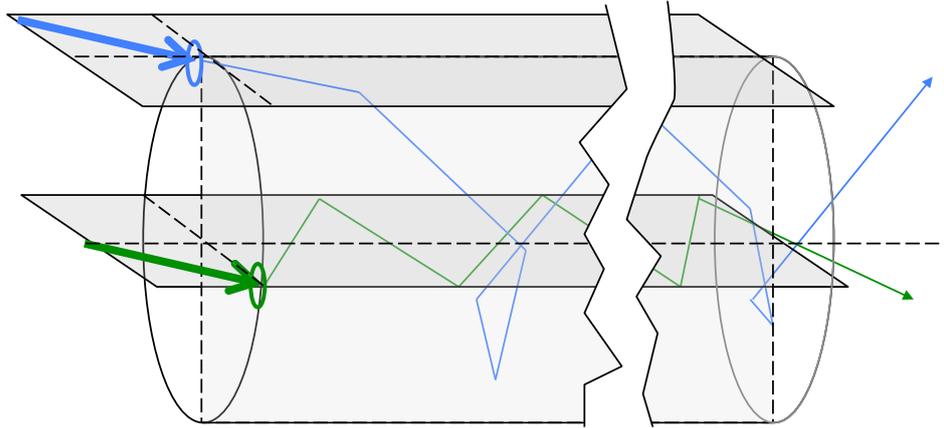

*Figure 14: Geometry of illumination to study the LNA. The illuminated spot on the fibre endface (grey) is shown red and the direction of the incident light is shown by the arrows which are contained within planes parallel to the horizontal meridian shown. Blue - 90 deg illumination azimuth where all rays are skew. Green - illumination azimuth where all rays are meridional (in the absence of FRD).*

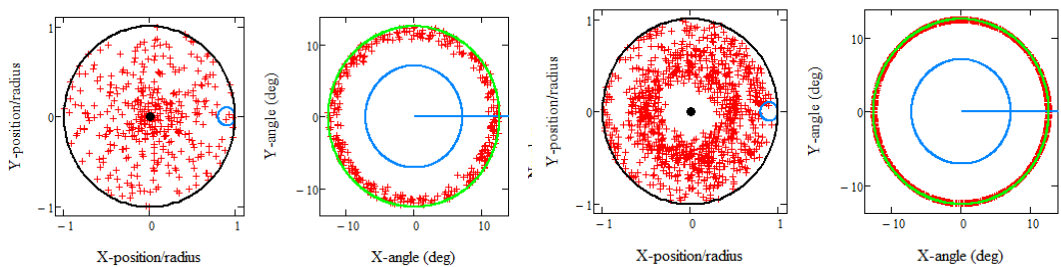

*Figure 15: Ray diagram for a collimated input beam at angle 12.7 deg. Each pair of plots consists of the nearfield (left) and farfield (right). The illuminated spot is shown by the small blue circle in the nearfield diagram. The azimuth direction of the incoming light is shown by the blue bar in the farfield diagram. In this case the illumination azimuth (defined in §6) is 0 deg. Left – with typical FRD (as defined in §4) included in the simulation. Note that the near-field pattern is peaked in the centre. Right – without FRD. The near-field pattern now shows a deficit of rays in the centre. In the farfield plots, the green circle (where visible) represents the locus of expected datapoints. The large blue circle represents the notional acceptance f/4.5 beam of the spectrograph.*

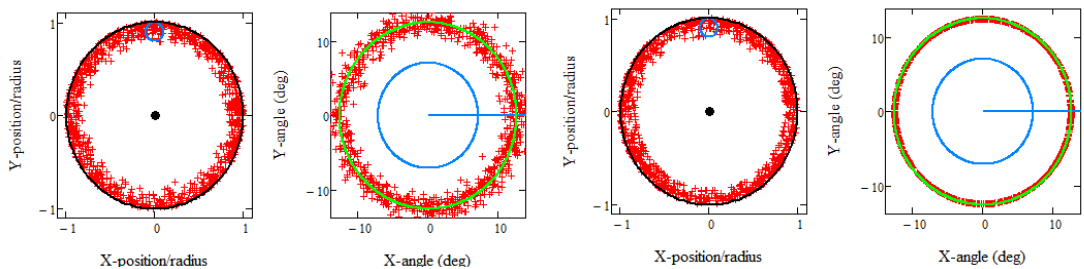

*Figure 16: Same as previous figure but for illumination azimuth 90 deg.*

The role of FRD in this process is interesting. When the illumination azimuth is 0, the rays are peaked towards the centre in the nearfield but when FRD is removed, there is a deficit of rays in the centre. Presumably, FRD has scattered the light enough to reduce the fraction of spiralling rays which would otherwise avoid the fibre axis. For larger azimuth angles, the same tendency to decrease the fraction of spiralling rays is seen when FRD is introduced but it cannot defeat the strong tendency for skew rays to dominate so that the LNA is defeated.

To examine this behaviour in more detail, we calculate the throughput as a function of the illumination angle (i.e. the input ray angle with respect to the optical axis) near the critical angle for different illumination azimuths with and without FRD (*Figure 17*). The simulations are for the case where the illumination azimuth is 0, 45 and 90 deg. When this angle is zero, the rays strike the core/cladding interface at the steepest angles within a meridian and so are most likely to exceed the critical angle, which in this example is 12.7 deg.

Without FRD, there is a very sharp cutoff as this angle is reached. When the illumination azimuth is varied so that rays strike the boundary at a lesser angle (45 and 90 deg), the cutoff is shifted to large illumination angles and becomes less distinct, showing a gradual rolloff rather than a sharp transition.

When FRD is added, the 0 degree azimuth rays now exhibit a rolloff instead of a sharp cutoff. Presumably this is because of the random perturbations introduced. The effect of FRD at the other azimuths (45 and 90 deg) is less marked because they already have had their arrival angles modified by the geometry of the interaction which generates significant numbers of skew rays in contrast to the purely meridional rays when the azimuth angle is zero.

From this we conclude that fibre output depends on the details of the illumination when close to the LNA. Although the case that we have examined in detail is extreme, it demonstrates that the output may be sensitive to variations in the input illumination whether by design or due to errors in manufacturer and alignment. In particular, illuminating fibres close to the LNA to reduce FRD (i.e. to minimise the fraction of rays exiting at large angles) is unlikely to be successful since many rays will be lost into the cladding and the angular distribution of the remaining rays will be hard to predict and sensitive to errors in alignment.

As a further check for less extreme cases which might be encountered in real situations, we simulate the effect of illuminating the fibre with a filled cone at F/2.5 equivalent to the LNA of typical fibres (0.22) with varying amounts of FRD (parameterised by $\Delta\theta_F$) for two cases: with the input of the fibre core uniformly illuminated and when illuminated by a spot of radius 0.6*R* offset by 0.4*R* so that it is just contained within the fibre core. The results are shown in ***Figure 18***.

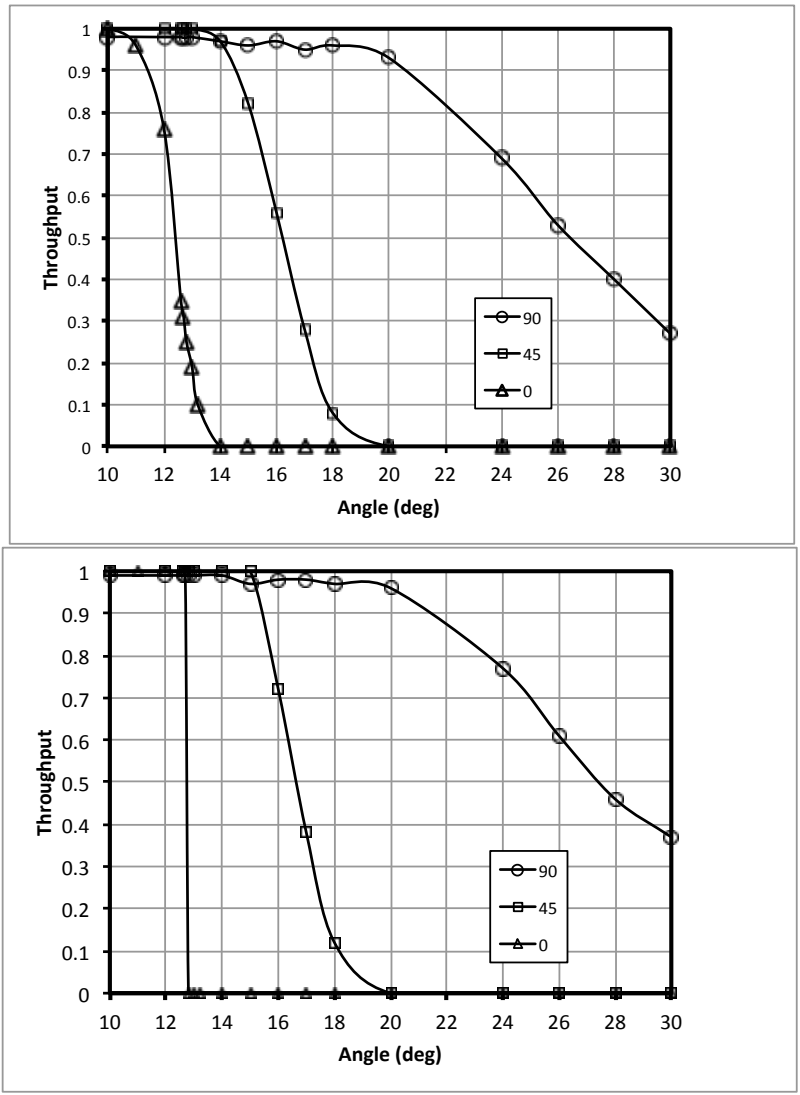

*Figure 17: Throughput as a function of the angle of illumination with respect to the fibre axis for different illumination azimuths as described in the text and shown in the legend. Top – with typical FRD. Bottom – witout FRD.*

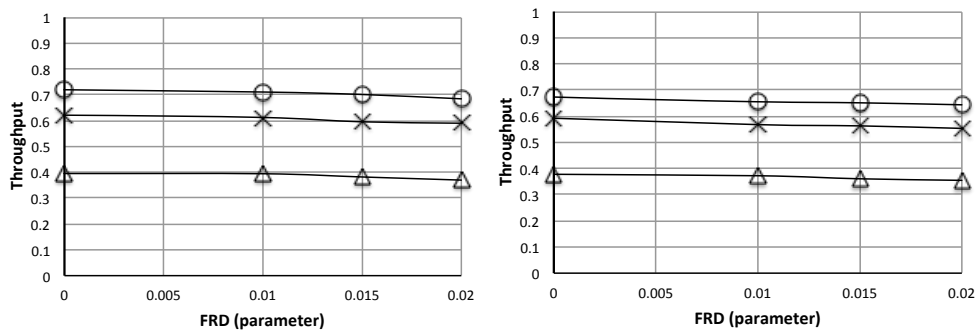

*Figure 18: Throughput of a fibre illuminated at F/2.5, close to the assumed LNA = 0.22 of a fibre. Left – for uniform illumination of the fibre. Right – for an offset spot as described in the text. The three curves are for an angular offset of the fibre with respect to the incoming illumination of 0, 5 and 10 deg from top to bottom respectively. The ordinate is the simulated amount of FRD parameterised by $\Delta\theta_F$. The error bars on the simulated data are smaller than the symbols.*

From this, it can be seen that varying the amount of FRD has little effect, but the system is fairly sensitive to the angular alignment of the fibres – an effect which increases slightlywith non-uniform illumination. This is expected from the simulation with collimated input light. Illumination with a filled cone might be expected to integrate over the illumination azimuth so as to cancel these effects out. But with non-uniform illumination, the cancellation may be imperfect. Indeed there is a residual effect, a drop of 4-8% of the remaining throughput.

We recommend that careful attention be taken of this effect in fibre systems operating close to the LNA, such as those deployed at prime focus where the input beam is very fast (F/2-F/3) especially if the fibres are tilted to acquire targets within their patrol field.

On the positive side, the simulations also suggest that the effect of the LNA could be reduced by deliberately increasing the population of skew rays. This could be done by illuminating the fibres via a spot offset from the fibre axis but implies very precise control of the input ray angles.

## 7   Non-standard fibres

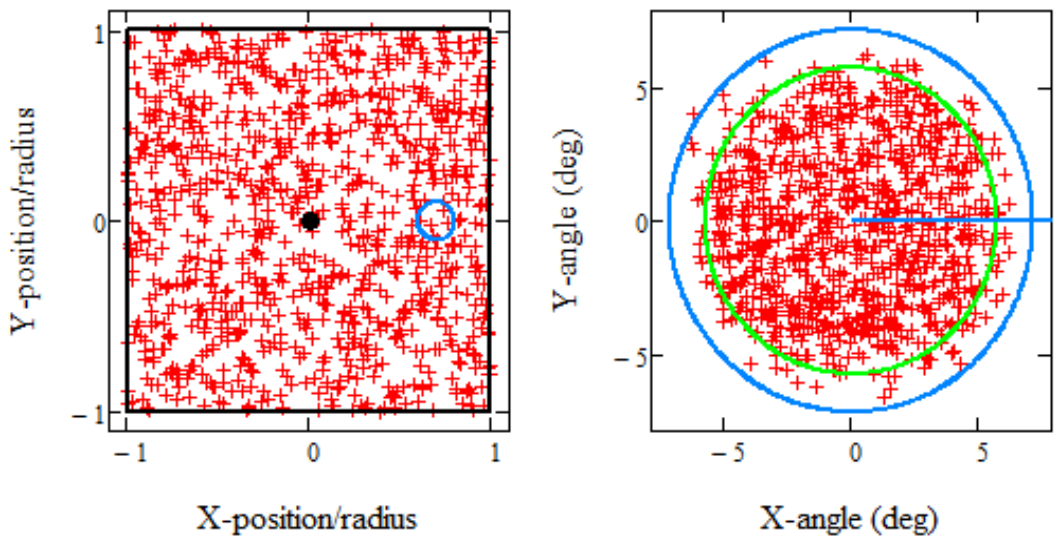

*Figure 19: Ray diagram for a square fibre illuminated through the spot indicated by the small circle in the nearfield (left) by a filled beam at F/5.*

Recently, it has been realised that square and octagonal fibres improve scrambling (e.g Avila et al. 2010, Chazelas et al 2010, Perruchot et al 2011). The aim of this section is to examine by simulation fibres which have a cross-section which is: (a) polygonal; (b) elliptical; and (c) tapered (vary linearly with position along the fibre axis). In particular, we wish to understand if there is likely to be any penalty in extra FRD because of the unusual geometry.

## 7.1 Polygonal cross-section

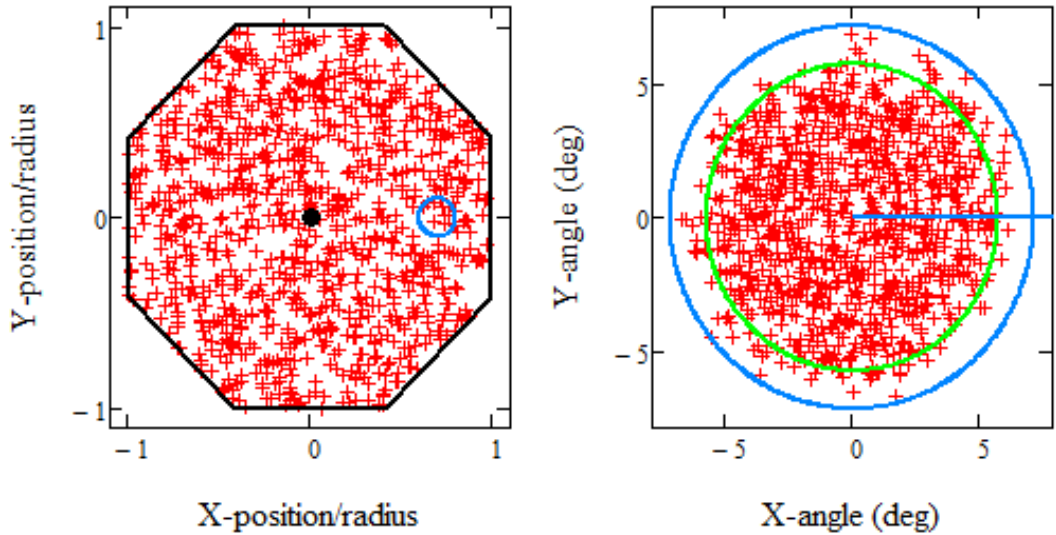

*Figure 20: Same as the previous figure but for an octagonal cross-section.*

*Figure 19* and *Figure 20* show ray diagrams for square and octagonal fibres where the input spot radius is 0.1*R* of the fibre radius offset by 0.8*R* (here R is the mean radius for octagonal fibres and edge-to-edge width for square fibres). It is evident (by tracing larger numbers of rays) that in both cases, the nearfield scrambling is very good. The farfield distributions are very similar to each other and to those of circular fibres (*Figure 3*).

We should not expect FRD to be increased by the shape of the aperture because meridional rays, in the absence of FRD, are reflected with fixed angle off the core-cladding boundary in the same way as for round fibres. (This effect should not be confused with the increase in incidence angle when light is fed into a fibre from *lenslets* with polygonal apertures. In that case, the lenslet aperture generates higher angles along symmetry axes - e.g. along the diagonals of a square lenslet. That effect is to increase the input angle, not the output angle for a constant input angle.[4])

Radically different behaviour is seen when the input is a collimated beam (*Figure 21*). Here the farfield is illuminated only at certain azimuths defined by the azimuth of the collimated beam and its reflection about the axes of symmetry of the polygon. If the FRD of a real fibre was assessed in this way, the results might seem puzzling. However it should nevertheless produce the correct result when the output is summed in azimuth (in this case the annulus has width 1.55 deg and a radius of 10.0 deg).

---

[4] However, for a lenslet-coupled fibre, the fibre+lenslet system will appear to generate extra FRD as etendue is conserved along the lenslet symmetry axes.

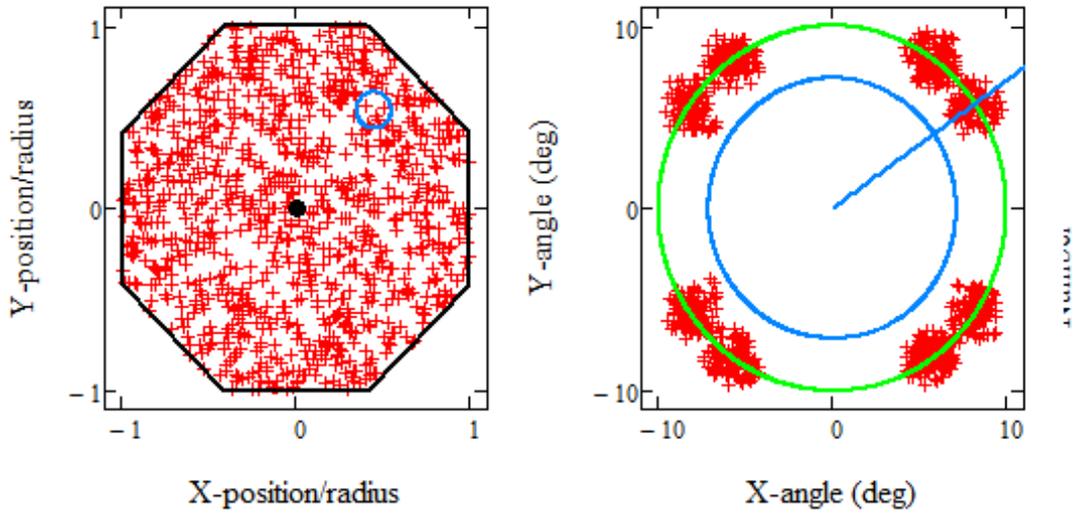

*Figure 21: ray diagram for a hexagonal fibre illuminated via the spot indicated in the nearfierld diagram (left) for a collimated beam of 10 deg oriented at the azimuth shown by the bar in the farfield diagram (right).*

We conclude therefore that we expect polygonal fibres to scramble better than circular ones, but should exhibit similar FRD to that of a circular fibre either in operation (with filled-cone input) or during test (with a collimated beam if azimuthally-averaged).

Although the previous publications on polygonal fibres quoted above have not provided clear results for FRD, recent tests show that some polygonal fibres have excellent FRD in line with these predictions (Murray, Moseley, and Allington-Smith 2012 in preparation; Poppett 2011).

## 7.2 Elliptical fibres

Multimode fibres with an elliptical cross-section are of interest in multiplexed spectroscopy because they potentially offer a way to image-slice in order to decrease the slitwidth (defined by the minor axis of the ellipse) while decoupling it from the size or shape of the aperture on the sky over which light is collected; so that the product of resolving power and throughput is increased. This could be achieved either by rotating the fibre input to align the ellipse with the projected orientation of an extended target; or, in conjunction with tapering (see below), to project a round aperture on the sky.

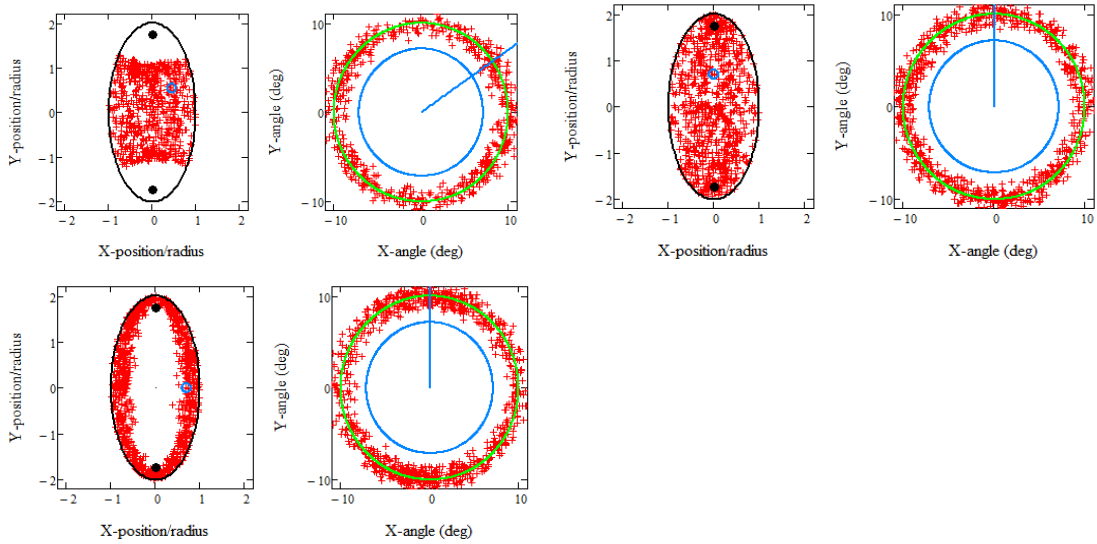

*Figure 22: As previous figure but for an elliptical fibre with cross-section shown in the near-field diagram. The foci are indicated by the black dots. The different illumination conditions are indicated by the illumination spot and illumination angle as marked.*

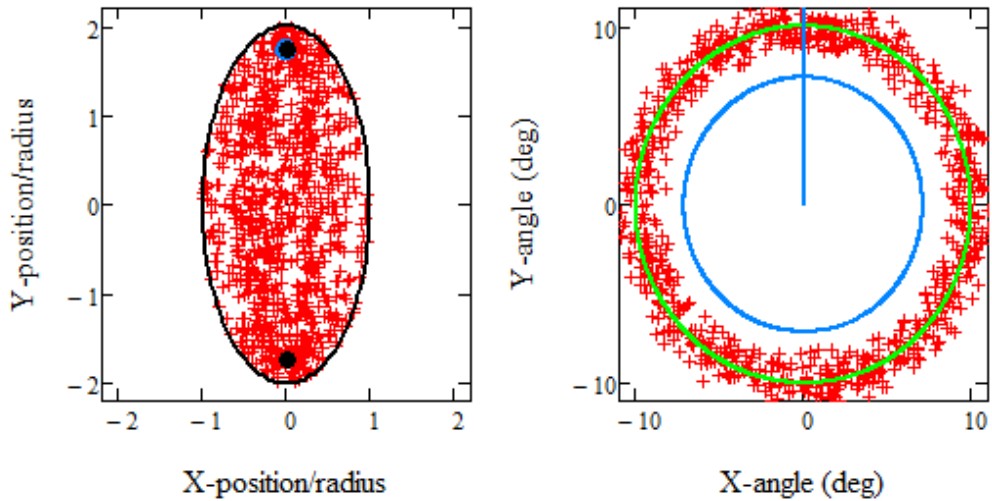

*Figure 23: As previous figure but with illumination spot located at the top focus.*

Figure 22 shows results when the fibre is illuminated by a collimated beam at an angle of 10 deg; the same configuration as in Figure 21. In contrast to that case, we find a uniform ring in the farfield but a radically different nearfield pattern which is sensitive to the way in which the fibre is illuminated.

For input in a filled cone over a large centrally-located spot, the nearfield pattern is quite uniform while the farfield output is similar to that from a circular fibre. Note that we should not expect an elliptical farfield pattern since Etendue is conserved (apart from FRD) so the input and output angles should be the same at all azimuths.

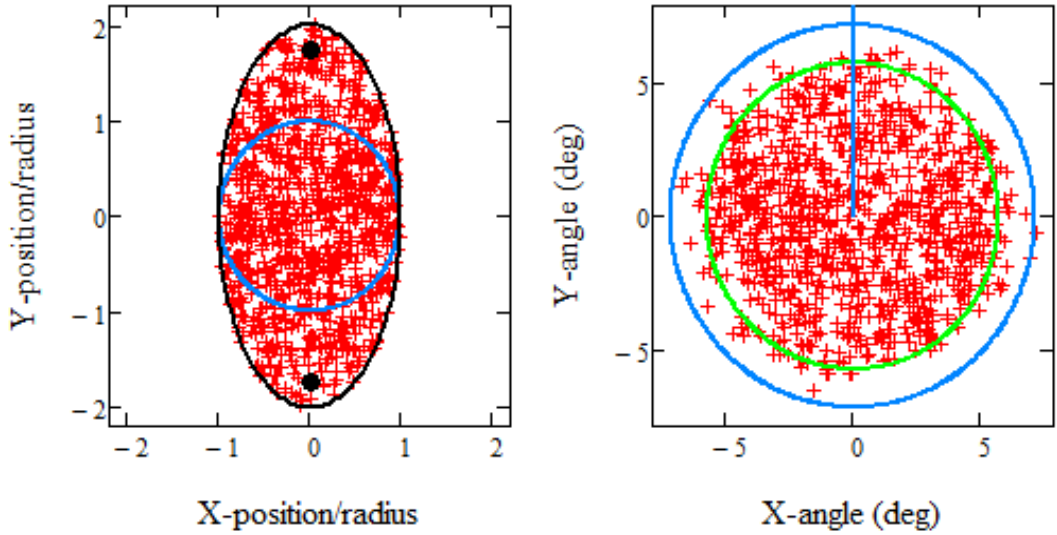

*Figure 24: Same as previous figure but for a filled F/5 cone and for a large centrally-located illumination spot.*

In principle we can exploit the geometry of an ellipse by injecting light at one focus and then see it emerge preferentially at the other. This is shown in *Figure 23* where the density of rays is highest at the two foci. This feature could be used to make the illumination at the slit more uniform to counteract the reduction in the illumination as the ellipse becomes narrow near the ends of the major axis.

## 7.3 Tapered fibres

The cross-section of fibres can be varied along their length to form a tapered fibre. One of the uses of this capability is to optimise light coupling between beams of different focal ratio. For example, a 2:1 taper could be used to couple an F/2 primary focus to an F/4 spectrograph. The fibre diameter would double in size from the focus to the spectrograph split as expected from conservation of Etendue, but the system will suffer from FRD and – most importantly – the incoming beam may approach or exceed the LNA.

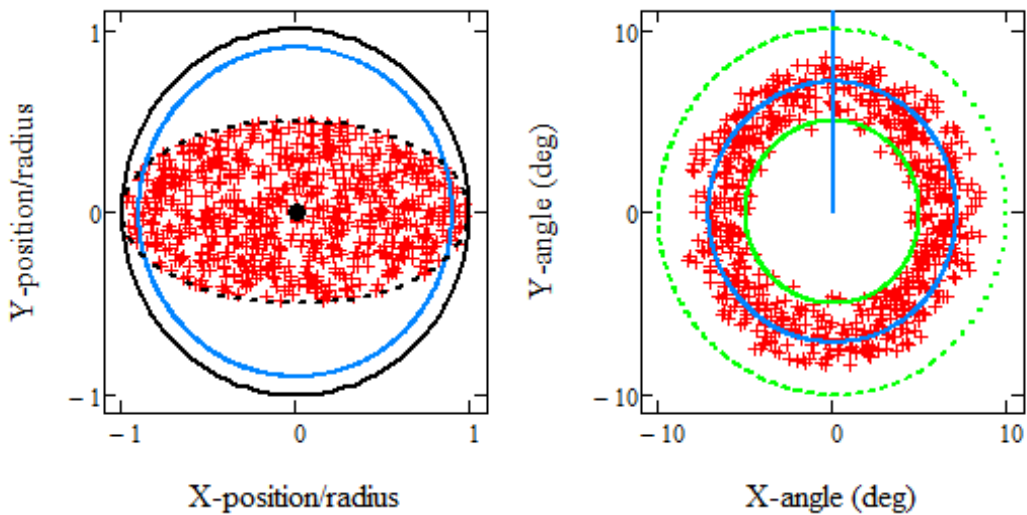

*Figure 25: Ray diagram for a fibre that tapers in the y-axis from the round aperture shown in the nearfield diagram (left) to the dotted ellipse at its end. The farfield diagram (right) shows the input angle of the collimated beam and (green-dashed) the angle expected from conservation of Etendue in the tapered axis.*

One option which may be of interest is a tapered elliptical fibre where the taper affects only one axis, so that the ellipticity changes along the fibre length. An example is simulated in *Figure 25* for a collimated beam injected at 5 deg. The output nearfield pattern in quite uniform, but the farfield patern shows a spread to larger angles than at the input. This is not surprising since Etendue conservation requires that light illuminating the minor-axis will exit at a higher angle. However, the surprise is that there is a lack of rays at large angles. The region between the two green circles should be uniformly populated with rays. In fact the throughput has dropped to 75%. For beams input at smaller angles, the missing rays are present as expected. In the case shown the maximum output ray angles should be 10 deg which is comfortably inside the LNA. Eliminating FRD and macro-bending did not improve the situation. An alternative view is that the LNA is decreased by the taper so that rays are more likely to exceed the critical angle, but this formulation requires that the effective speed of the exiting light is constant. In other words, we must not double-count the effect of etendue conservation and LNA reduction: they are the same thing; hence the loss of throughput is surprising.

Note that, in some cases the rays may travel backwards as decreasing diameter increases the ray angles to zero and then to negative values corresponding to travel back to the input so that the light is not transmitted through the fibre. However this seems contrived since one would expect the light to exceed the critical angle at some point and so exit from the fibre core into the cladding. Only in unusual circumstances could that barrier be stepped over to leave a ray free to follow its inverted career.

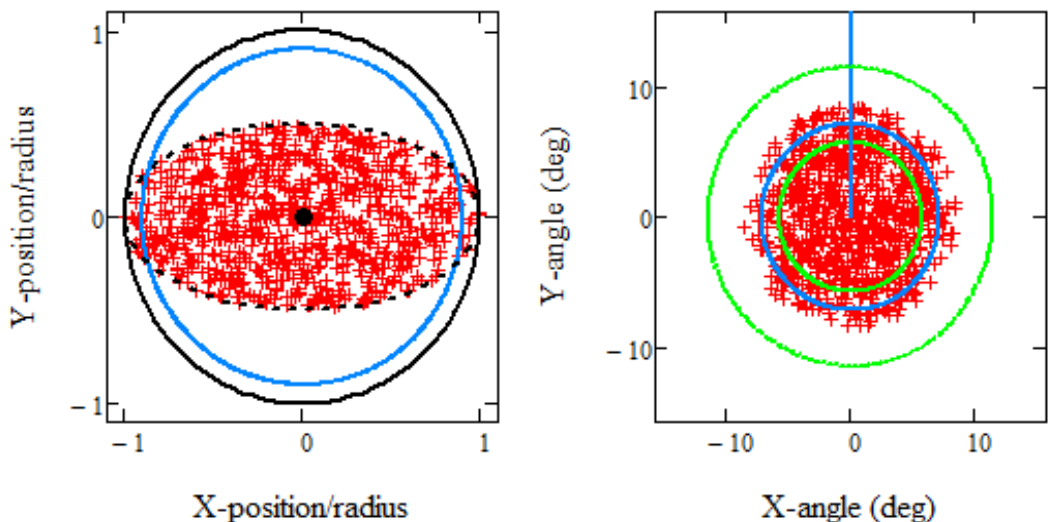

*Figure 26: Same as previous figure but for an F/5 filled-cone of illumination.*

For a filled cone, the result (*Figure 26*) is similar to that of untapered circular fibres with a circular top-hat farfield distribution of width intermediate between the input F/5 and F/2.5. However the throughput is only 89% suggesting that some rays were lost by exceeding the Numerical Aperture as discussed above.

# 8 Conclusions

We have shown that it is possible to simulate the main characteristics of optical fibres (macro- and micro-bending) in the extreme multimode limit by ray-tracing with appropriate perturbations of the angle of reflection at the core-cladding boundary. This may provide a convenient way to explore the properties of different fibre configurations without extensive experimental work. However, some features of fibres are not within the scope of the model. In particular, the contribution to FRD of diffraction is not accounted for and geometric effects due to lenslet coupling are not included in the model in its present form. Similarly, coherence effects are not presently included which will reduce the simulation accuracy for thin fibres or where the number of supported modes is small. Experimental work will still be required to account for these effects and provide calibration of the model.

This technique also allows exploration of the properties of non-standard fibres (polygonal cross-section, tapered etc.) including some not yet available. We can also identify configurations where the performance may be sensitive to details of the fibre illumination. In these configurations a real system might show unexpected variations from fibre to fibre and from time to time leading to a degraded SNR. The uniformity of SNR is an issue for the new generation of fibre instrumentation addressing basic cosmological questions such as the dark energy.

Using the model, we find that scrambling primarily affects the shape of the near-field pattern but has negligible effect on the barycentre. FRD helps to homogenise the nearfield pattern but does not make it completely uniform. Fibres with polygonal cross-section improve scrambling without amplifying the FRD in contrast to some expectations.

Elliptical fibres, in conjunction with anamorphic tapering may offer an efficient means of image slicing to improve the product of resolving power and throughput but the result is sensitive to the details of illumination.

## Acknowledgements

We thank the Science and Technology Facilities Agency for financial support. We also thank Jerry Edelstein, Bruno Chazelas, Claire Poppett and Ben Moseley for helpful discussions and encouragement .